\documentclass[aps,prl,twocolumn,groupedaddress,showpacs,superscriptaddress,nobibnotes]{revtex4}
\usepackage{graphicx}
\usepackage{epsfig}

\begin{document}

\title{Magnetic and Electric Excitations in Split Ring Resonators
}

\author{Jiangfeng Zhou}%

\address{Department of
Electrical and Computer Engineering and Microelectronics Research
Center, Iowa State University, Ames, Iowa 50011}%
\address{Ames Laboratory and Department of Physics and Astronomy,Iowa
State University, Ames, Iowa 50011}%

\author{Thomas Koschny}
\address{Ames Laboratory and Department of Physics and Astronomy,Iowa State University, Ames, Iowa 50011}
\address{Institute of Electronic Structure and Laser - FORTH,and
Department of Materials Science and Technology, University of Crete,
Greece}

\author{Costas M. Soukoulis}
\address{Ames Laboratory and Department of Physics and
Astronomy,Iowa State University, Ames, Iowa 50011}
\address{Institute of Electronic Structure and Laser - FORTH,and
Department of Materials Science and Technology, University of Crete,
Greece}

\begin{abstract}
We studied the electric and magnetic resonance of U-shaped SRRs. We
showed that higher order excitation modes exist in both of the
electric and magnetic resonances. The nodes in the current
distribution were found for all the resonance modes. It turns out
that the magnetic resonances are the modes with odd-number of
half-wavelength of the current wave, i.e. $\lambda/2$, $3\lambda/2$
and $5\lambda/2$ modes, and the electric resonances are modes with
integer number of whole-wavelength of current wave, i.e. $\lambda$,
$2\lambda$ and $3\lambda$ modes. We discussed the electric moment
and magnetic moment of the electric and magnetic resonances, and
their dependence to the length of two parallel side arms. We show
that the magnetic moment of magnetic resonance vanishes as the
length side arms of the SRR reduces to zero, i.e. a rod does not
give any magnetic moment or magnetic resonance.
\end{abstract}


\maketitle

\section{Introduction}
The idea of negative index materials (NIMs), i.e. materials with
both negative electrical permittivity, $\epsilon$, and magnetic
permeability, $\mu$, was first introduced by Veselago
\cite{Veselago_1968}. However, it was only recently that such
materials were investigated experimentally,
\cite{SCI_smith_pendry_review_2004,Adv_Mat_soukoulis_review_2006,
Linden_IEEE_QE_2006,Nat_Phonotics_shalaev_2007,Sci_soukoulis_2007}.
Although it has been well known how to obtain $\epsilon<0$ material
easily (e.g. using lattice of metallic wires), the realization of
$\mu<0$ (especially at high frequencies) response had been a
challenge, due to the absence of naturally occurring magnetic
materials with negative $\mu$. In 1999, Pendry et al.
\cite{IEEE_pendry_srr} suggested a design made a two concentric
metallic rings with gaps, called split ring resonators (SRRs), which
exhibit a $\mu<0$ around the magnetic resonance frequency
$\omega_m$. Immediately after Smith
 et al. \cite{PRL_Smith_first_NIM} fabricated the first negative
 index material at GHz frequencies. Recently different groups
 observed \cite{sci_yen_Zhang_Smith_2004,science_Linden_100THz,PRL_Enkrich_usrr,katsarakis_OL_2005}
 indirectly negative $\mu$ at THz frequencies. In most of the THz
 experiments, only one layer of SRRs were fabricated on a substrate
 and the transmission, $T$, was measured only for propagation
 perpendicular to the plane of the SRRs, exploiting the coupling of
 the electric field to the magnetic resonance of the SRR via
 asymmetry \cite{katsarakis_apl_2004}. It was realized that one
 only need the single SRR to see the magnetic resonance effects.
 This way is not possible to
 drive the magnetic permeability negative. One reason is that is
 very difficult to measure with the existing topology of SRRs and
 continuous wires both the transmission, $T$, and reflection, $R$,
 along the direction parallel to the plane of the SRRs. So there is
 a need for alternative, improved and simplified designs that can be
 easily fabricated and experimentally characterized. This new design was
 recently achieved in the GHz region \cite{PRB_zhou_cwp,
 APL_zhou_cwp} and the THz region
 \cite{PRL_zhang_2005_037402,PRL_zhang_2005_137404,dolling_OL_2006,science_soukoulis_2006}
  by the use of finite
 length of wires and the fishnet topology. Very recent work has
 moved the negative refractive index into optical wavelength
 \cite{dolling_fishnet_780nm_2007,Opl_Chettiar_2007}.

 In this manuscript we systematically studied the electric and
 the magnetic resonance response of U-shaped SRRs for different
 propagating directions. The effective electric permittivity,
 $\epsilon$, as well as the magnetic permiability, $\mu$ will be
 extracted by the retrieval procedure
 \cite{PRB_smith_retrieval_2002,PRE_Smith_retrieval_2005,PRB_koschny_retrieval_2005}.
 In addition, the current distribution along the sides of the U-shaped
 SRR will be numerically calculated. We show that the
 magnetic resonance are the modes with odd-number of
 half-wavelength of the current density wave, while the electric resonance are modes with
 integer number of whole-wavelength the current density wave. In addition we studied the
 dependence of the electric and magnetic resonance as a function of
 the length of the side arms of the U-shaped SRR. It is found that
 the magnetic moment of the U-shaped SRR vanishes as the length of side
 arms of SRR reduces to zero. So there is no magnetic moment or
 magnetic resonance for a metallic rod. One needs the side arms of the U-shaped SRR in
 order to have a magnetic moment.
%

\section{Electric and Magnetic Responses of SRRs}
A common constituent to provide magnetic response in metamaterials
is the Split-Ring resonator (SRR). The SRR in its simplest form
consists of a highly conductive metallic ring which is broken in one
(or several) location(s) by a non-conductive gap of air or other
dielectric materials. If this ring is placed in a temporally varying
magnetic field an electric circular current is induced in the
metallic ring which in turn leads to charge accumulating across the
gaps. The electric field which builds due to the charge at the gap
counteracts the circular current leading to energy stored
(predominantly) in vicinity of the gaps and magnetic field energy
concentrated in the region enclosed by the ring. The SRR is thus a
resonator which couples to a perpendicular magnetic field and can be
characterized by the effective capacitance of the gaps and effective
inductance of the loop define by the ring. It can be understood in
terms of a resonant $LC$ circuit with a resonance frequency
$\omega_m^2=1/LC$, where $L$ is the inductance and $C$ is the
capacitance of the SRR. The resonant response of the circular
current in the SRR to an external magnetic field leads to a resonant
magnetic moment which may reach large negative values for array of
SRRs such that the size of the SRR is much smaller than the
wavelength of an incident electromagnetic wave around the resonance
frequency behaves as a homogeneous effective medium with at negative
(resonant) permittivity $\mu_\mathrm{eff}(\omega)$.

In this letter we report numerical results of electric and magnetic
responses of a periodic lattice of SRRs for different orientations
of the SRRs with respect to the external electric field, $\vec{E}$,
and the direction of propagation, $\vec{k}$.

\begin{figure}[htb]\centering
  \includegraphics[width=8cm]{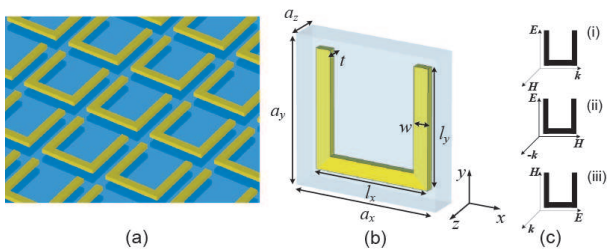}\\
  \caption{(a) Schematic of periodically arranged U-shaped SRR
  arrays; (b) a single unit cell with geometrical parameters; (c) three different
  configurations of incident electric field, $\vec{E}$, magnetic field,
  $\vec{H}$, and wave vector, $\vec{k}$.
  \label{fig1_structure}}
\end{figure}

Our numerical simulations were done with CST Microwave Studio
(Computer Simulation Technology GmbH, Darmstadt, Germany), which
uses a finite-integration technique, and Comsol Multiphysics, which
uses a frequency domain finite element method. The schematic of the
periodic U-shaped SRRs arrays and the geometry of a single unit cell
used in our numerical simulation were shown in Fig.
\ref{fig1_structure}(a) and (b), respectively. The SRRs exhibit
different responses to the incident electromagnetic (EM) wave with
respect to different configurations of incident electric field,
$\vec{E}$, magnetic field, $\vec{H}$, and wave vector, $k$, as shown
in Fig. \ref{fig1_structure}(c). First, it is well know that the
incident EM wave excites a magnetic resonance at $\omega_m$
\cite{IEEE_pendry_srr,SCI_smith_pendry_review_2004,Adv_Mat_soukoulis_review_2006,
Linden_IEEE_QE_2006,Nat_Phonotics_shalaev_2007,Sci_soukoulis_2007}
if the external magnetic field, $\vec{H}$, is perpendicular to the
SRR plane and the wave vector, $\vec{k}$, is parallel to the SRR
plane (Fig. \ref{fig1_structure}(c.i)). Second, farther study
reveals that the incident EM wave with $\vec{k}$ perpendicular to
the SRR plane and $\vec{E}$ parallel to the bottom part of SRR (Fig.
\ref{fig1_structure}(c.ii)) also excites a magnetic resonance at
$\omega_m$ \cite{katsarakis_apl_2004}. This magnetic resonance
results from a circular current induced by the external electric
field, $\vec{E}$, because of the asymmetry of the SRR in the
direction of $\vec{E}$, and therefore was called the electric
excitation coupling to the magnetic resonance (EEMR). The EEMR is
very valuable for the experimental demonstration of the magnetic
resonance at optic frequencies, because it is very difficult to
measure the transmission and reflection with the incident EM wave
parallel to the SRR plane at high frequencies
\cite{sci_yen_Zhang_Smith_2004,
science_Linden_100THz,PRL_Enkrich_usrr,katsarakis_OL_2005,katsarakis_apl_2004}.
Third, the SRRs also exhibit a short-wire-like electric resonance
\cite{PRL_koschny_2004} at $\omega_0$ with the incident electric
field, $\vec{E}$, parallel to the side part of SRRs (Fig.
\ref{fig1_structure}(c.iii)). The electric resonance frequency
$\omega_0$ depends on the length of the side part of SRRs, $l_y$,
being higher for shorter $l_y$.

\begin{figure}[htb]\centering
  \includegraphics[width=6cm]{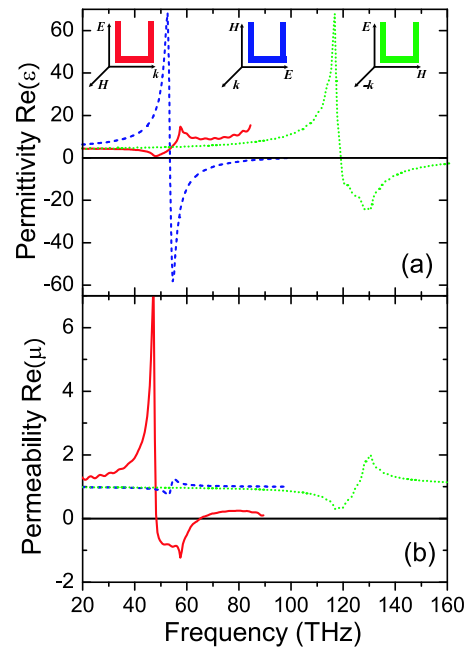}\\
  \caption{Extracted effective permittivity $\mathrm{Re}(\epsilon(\omega))$ (a)
  and effective permeability $\mathrm{Re}(\mu(\omega))$ (b) for the magnetic resonance (red solid),
  the electric excitation coupling to the magnetic resonance (EEMR) (blue dashed) and the short-wire-like
  resonance (green dotted) of  USRRs. The geometric parameters are
  $a_x=a_y=1\ \mu$m, $a_z=0.2\ \mu$m, $l_x=l_y=0.8\ \mu$m, $w=0.1\ \mu$m and $t=0.05\ \mu$m.
  \label{fig2_three_polization}}
\end{figure}

Using the retrieval procedure
\cite{PRB_smith_retrieval_2002,PRE_Smith_retrieval_2005,PRB_koschny_retrieval_2005},
we calculated the effective permittivity $\epsilon(\omega)$ and
permeability $\mu(\omega)$, both real and imaginary part, from the
simulated transmission,$T$, and reflection, $R$. Fig.
\ref{fig2_three_polization} shows the extracted real part of the
effective permittivity $\mathrm{Re}(\epsilon(\omega))$ and
permeability $\mathrm{Re}(\mu(\omega))$ for three different
polarized incident EM waves as shown in Fig.
\ref{fig1_structure}(c). As expected, the magnetic resonance,
measured by $\mu$ (red solid)), occurs at around $\omega_m=55$ THz.
The EEMR gives a similar resonance at roughly the same frequency,
shown as a resonance shape in $\epsilon$ (blue dashed), which
indicate the response is due to the electric coupling of the
incident EM wave to SRR. At a higher frequency, $\omega_0=135$ THz,
the short-wire-like electric resonance of the SRR occurs, given by
the resonance behavior of $\epsilon$ (green dotted). Due to the
periodicity effect, whenever a magnetic resonance occurs in
$\mathrm{Re}(\mu)$, an electric anti-resonance will be seen in
$\mathrm{Re}(\epsilon)$ simultaneously and vice versa
\cite{PRB_koschny_retrieval_2005,PRE_koschny_retrieval_2003}. The
periodicity effect becomes more noticeable as the wavelength of EM
wave is comparable with the lattice constant of the SRR lattice. For
the magnetic resonance (red solid), the lattice constant, $a_x=1 \
\mu$m, in the propagating direction along the wave vector,
$\vec{k}$, is comparable to the resonance wavelength,
$\lambda_m=5.45\ \mu$m, hence, a strong distortion in the negative
part of resonance in $\mathrm{Re}(\mu)$ and a significant
anti-resonance in $\mathrm{Re}(\epsilon)$ were observed. On the
other hand, for the EEMR, the lattice constant, $a_x=0.2\ \mu$m, is
much smaller than $\lambda_m$, so the periodicity effect is much
weaker and therefore a sharp resonance in $\mathrm{Re}(\epsilon)$
(blue dashed) with a weak anti-resonance in $\mathrm{Re}(\mu)$
occurs.

\section{Higher Order Excitation Modes of SRRs}
\begin{figure}[htb]\centering
  \includegraphics[width=8cm]{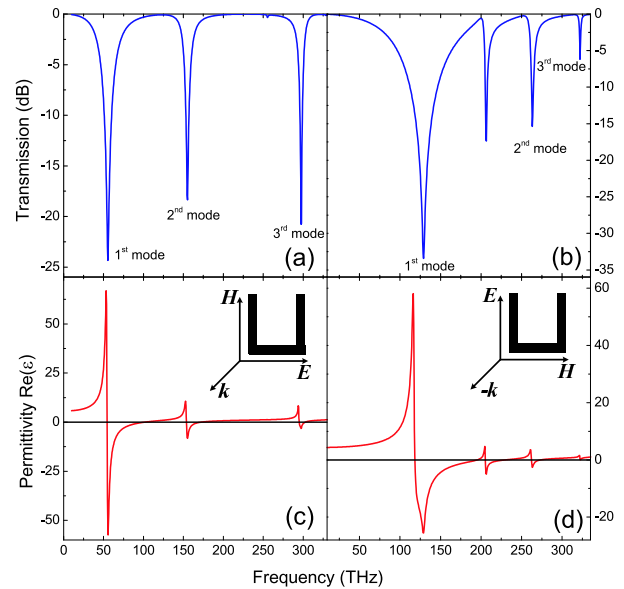}\\
  \caption{Transmission spectra (a,b) and the extracted permittivity, $\mathrm{Re}(\epsilon)$, (c,d)
  of the U-shaped SRRs response to the incident EM wave. The directions of $\vec{E}$,
  $\vec{H}$ and $\vec{k}$ were shown in the insets.\label{fig3_T_eps_ly_0.8}}
\end{figure}

Beside the typical electric and magnetic response, the SRRs also
exhibits higher order excitation modes. As shown in Fig.
\ref{fig3_T_eps_ly_0.8}(a), for the incident EM wave with the
propagating direction perpendicular to the plane of SRRs, we found
three successive dips at 55 THz, 155 THz and 300 THz in the
transmission spectra. Correspondingly, at the same frequencies, we
found three resonance shapes in $\mathrm{Re}(\epsilon)$ (Fig.
\ref{fig3_T_eps_ly_0.8}(c)). The first excitation mode occurs at the
lowest frequency $f_{m0}=55$ THz, which is the usual EEMR response,
and has the strongest resonance in $\mathrm{Re}(\epsilon)$. The
magnitudes of resonance of the second and third excitation modes,
occur approximately at 3$f_{m0}$ and 6$f_{m0}$ respectively, and are
much weaker then the first excitation mode. Higher order excitation
modes of the short-wire-like response also been observed (Fig.
\ref{fig3_T_eps_ly_0.8}(b) and (d)). In Fig. \ref{fig3_T_eps_ly_0.8}
(b), four dips were found in the transmission spectra, which result
from the first, second and third order electric excitation modes of
SRRs. Detailed studies show that the second (205 THz) and the third
(265 THz) modes have similar current density distribution, so they
are both considered as the second excitation mode.

As a very crude picture we could imagine the resonances of the SRRs
as charge density waves on a rod of a length equal to the arc length
of the SRR ring. This rod supports plasmonic modes
\cite{opl_rockstuhl_2006} (plasmonic means in this situation that
the inductance is coming from the electron mass and the capacitance
from the external electric field over the surface of the rod) which
occur at discrete frequencies whenever we have current nodes at the
ends of the rod. This picture is of limited value for the following
reasons: (i) the EM response of the modes, especially the
classification of electric vs.\@ magnetic response, depends on the
geometry and is entirely different for a U-shaped SRR than for a
straight rod although the qualitative current distribution over the
arc length is equivalent; (ii) the plasmonic dispersion depends on
the geometric inductance, i.e.\@ the energy stored in the external
field outside the metal, which is much larger for the SRR than for
the rod and spatially non-uniform along the SRR ring; and (iii) the
modes of coupling are different. Nevertheless, considering the
current distribution is an essential tool for understanding the
resonant modes of a SRR at higher THz frequencies. In the microwave
region only the remnants of the lowest order mode survive.

\begin{figure}[htb]\centering
  \includegraphics[width=9cm]{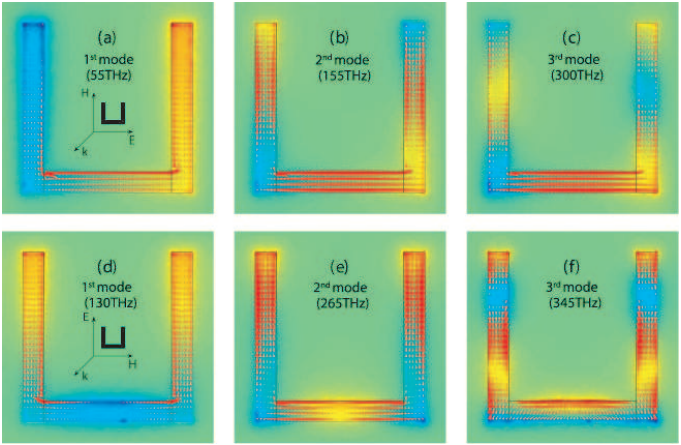}\\
  \caption{
 Distribution of the perpendicular component of the surface electric field
 (color scale; red positive, blue negative) and the bulk current density
 (arrows) for the lowest few resonant modes of the SRR.  The SRR metal is made
 of a Drude model for gold ($f_p=2175$ THz,
 $f_\tau=6.5$ THz), the geometry parameters are:
 $a_x=a_y=1\ \mathrm{\mu m}$, $a_z=200\ \mathrm{nm}$ (unit cell size),
 $l_x=l_y=800\ \mathrm{nm}$ (arm length), $w=100\ \mathrm{nm}$, $t=50\ \mathrm{nm}$
 (ring width and thickness, respectively).
 The current distributions are shown temporally $\pi/2$ phase shifted against
the charge distribution. \label{fig4_charge_current}}
\end{figure}

\begin{figure}[htb]\centering
  \includegraphics[width=6cm]{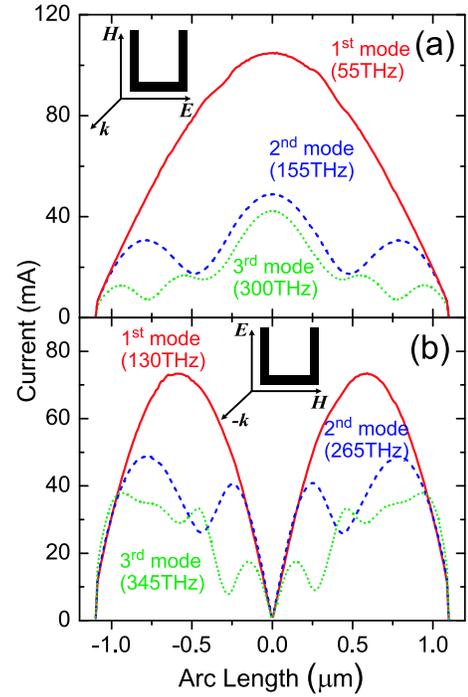}\\
  \caption{Current distribution of the lowest 3 modes.
  (a) $\lambda/2$, $3\lambda/2$, and $5\lambda/2$ mode for the EEMR
  response;
  (b) $\lambda/2$, $3\lambda/2$, and $5\lambda/2$ mode for the short-wire-like
  response. Due to the nonzero response of lower order modes, nodes of higher
  order modes only reach zero at the positions of the lowest nodes.
  \label{fig5_current}}
\end{figure}

Fig. \ref{fig4_charge_current} shows the distribution of the current
and charge density for the lowest three resonant modes of a U-shaped
SRR of side length, $l_y=800$ nm. The current density is obtained
directly from the simulation results, the charge density relates to
the perpendicular electric field at the surface of the metal. All
fields are time harmonic; the current distributions are shown
temporally $\pi/2$ phase shifted against the charge distribution.
The first three panels (a,b,c) of Fig. \ref{fig4_charge_current}
show the lowest three EEMR resonances for normal incidence to the
SRR with the electric field breaking the symmetry of the SRR and
thus we have coupling to the "magnetic"
resonance\cite{katsarakis_apl_2004}. The surface electric field
distribution is qualitatively equivalent to the pure magnetic
coupling (i.e.\@ propagation in the SRR plane with perpendicular
magnetic field), which case is however hard to realize
experimentally at such high THz frequencies.
All three modes have non-zero magnetic moment coming from all three
"arms" of the SRR. The number of current nodes (where charge
accumulates) increases with the resonance frequency from two
(fundamental mode, corresponding to the resonance of the effective
$LC$ circuit), one on either side of the "gap", to four and six,
which have additional nodes inside the continuous metal. The
electric excitation of these modes occurs via the polarization of
the bottom arm of the SRR by the electric field of the incident EM
wave. For a straight rod, these SRR modes would correspond to the
$\lambda/2$, $3\lambda/2$, and $5\lambda/2$ mode (Fig.
\ref{fig5_current}(a)); however, for the metallic rod, there is no
magnetic moment associated with the excitation modes.
Note that these modes also possess an electric dipole moment;
therefore the SRR has a combined magnetic and electric response in
this configuration.
For the other polarization, normal incidence to the SRR with the
electric field along the symmetry axis of the SRR, shown in panels
(d,e,f) of Fig. \ref{fig4_charge_current}, we see the analog
plasmonic modes with three, five and seven current nodes; all of
which reflect the mirror symmetry of the SRR and can thus have no
magnetic moment. For the metallic rod, they would correspond to the
even, $\lambda$, $2\lambda$, and $3\lambda$ modes (which cannot be
exited for the rod because of their vanishing dipole moment). In the
case of the SRR they do possess electric dipole moment in the $\bf
E$-direction, i.e.\@ along the two parallel arms of the SRR, and
represent a purely electric response of the SRR.


Fig.~\ref{fig5_current}(a) and (b) shows the distribution of the
total current (current density $\bf j$ integrated over the
cross-section of the SRR ring) over the arc length around the SRR
ring for the three resonant modes of
Fig.~\ref{fig4_charge_current}(a,b,c) and
Fig.~\ref{fig4_charge_current}(d,e,f) respectively. We clearly see
the different number of current nodes, which do not reach zero (for
the higher modes) because of the superposition with the non-resonant
response of the lower order modes (which are fairly broad due to the
high losses).  Also note the non-uniform spacing of the nodes for
the higher orders which are different from the straight rod and are
due to the curvature of the SRR and the coupling to the bottom arm
or side arms only.

\section{Electric and Magnetic Moments of SRRs}
We also studied the electric moment and the magnetic moment for the
electric excitation of the magnetic resonance (EEMR) and the
short-wire-like resonance of the U-shaped SRRs with different length
of two parallel side arms, $l_y$. The electric moment is calculated
by the integration of the subtraction between the electric
displacement $\vec{D}$ and vacuum electric displacement $\epsilon_0
\vec{E}$ over the volume of the whole unit cells.
\[
\vec{p}=\int{(\vec{D}-\epsilon_0 \vec{E})}\mathrm{d} \vec{r}
\]
\begin{figure}[htb]\centering
  \includegraphics[width=6cm]{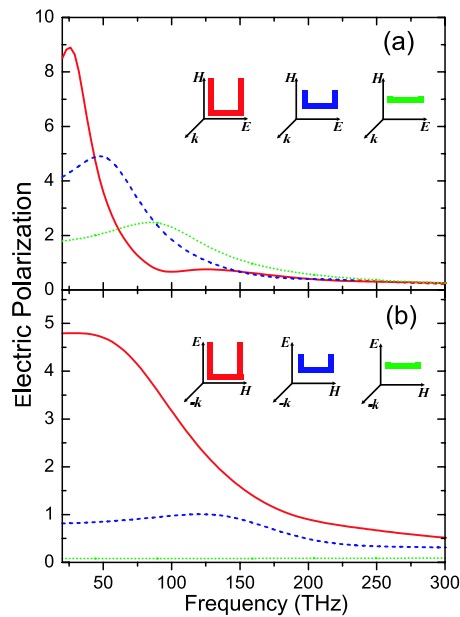}\\
  \caption{Magnitude of the normalized polarization density, $P$,
  of
  the U-shaped SRRs with $l_y=0.8\ \mu$m  (red solid), 0.4 $\mu$m (blue dashed)
   and 0.11 $\mu$m (green dotted), respectively.
  (a) $P_x$ component as $\vec{E}$ parallel to the bottom part of
  SRRs. The other two components $P_y$ and $P_z$ are nearly zero;
  (b) $P_y$ component as $\vec{E}$ parallel to the side part of
  SRRs. The other two components $P_x$ and $P_z$ are nearly zero.
  The polarizations of the incident EM wave are shown as the
  insets in the panel (a) and (b).
  \label{fig6_P}}
\end{figure}
Then the electric polarization, normalized by incident electric
field, $\bf E_0$, $\vec{P}=\vec{p}/(V E_0)$ is shown in Fig.
\ref{fig6_P}(a) and (b) for the EEMR and the short-wire-like
response of the SRRs respectively. It is clearly seen that the
lowest mode of resonance has the strongest electric moment although
the higher order excitation also has nonzero value. In Fig.
\ref{fig6_P}(a), the electric polarization, $P_x$, decreases as the
length of two parallel arms, $l_y$, decreases, because the EEMR
response becomes weaker and finally disappeared as the length $l_y$
close to zero (green dotted curve). However, the electric
polarization, $P_x$, does not vanish in the limiting case of
$l_y=0$, (green dotted curve), because the EEMR degenerates to the
short-wire-like electric resonance of the bottom arm of the SRR
which results in the non-zero electric polarization.
The polarization density along the y-direction, $P_y$, of the
short-wire-like resonance of the SRR (Fig. \ref{fig6_P}(b)) with the
incident electric field $\vec E$ parallel to the side arms of the
SRR, also decreases and shifts to higher frequencies as $l_y$
decreases.

The magnetic moment is calculated by employing the formula:
\[
\vec{m}=\frac{j\omega}{2}\int{\vec{r} \times (\vec{D}-\epsilon_0
\vec{E})}\mathrm{d} \vec{r}
\]

\begin{figure}[htb]\centering
  \includegraphics[width=6cm]{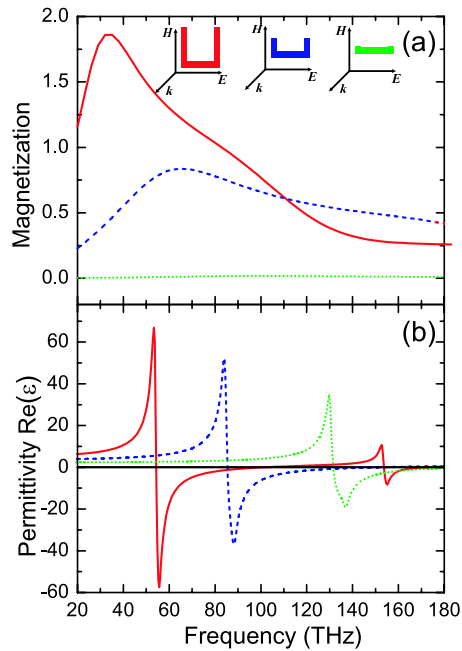}\\
  \caption{(a) Magnitude of the normalized magnetization, $M_z$, and (b) the
  extracted permittivity, $\mathrm{Re}(\epsilon)$, of the U-shaped SRRs with the length $l_y=0.8\ \mu$m
  (red solid), 0.4 $\mu$m (blue dashed) and 0.11 $\mu$m (green dotted), respectively. The polarizations of the incident EM wave are shown as the
  insets in the panel (a).
  The short-wire-like resonance with incident EM wave polarized as shown
  in Fig.~\ref{fig1_structure}(c.ii) has zero magnetic moment, and therefore is not shown here.
  \label{fig7_Mz}}
\end{figure}

In Fig.~\ref{fig7_Mz}(a), we show the magnetic moment, normalized by
the magnetic field of the incident EM wave, $\bf H_0$,
$\vec{M_z}=\vec{m_z}/(V H_0)$,  and in Fig.~\ref{fig7_Mz}(b), the
extracted permittivity, $\mathrm{Re}(\epsilon)$, for the magnetic
resonant modes of the SRR as a function of frequency for three
different U-shaped SRRs with different lengths of the two parallel
arms, $l_y$. It is clearly seen that the lowest order magnetic
resonance provides the strongest magnetic response; but also the
higher modes have non-zero magnetic moment. As one expects, reducing
the length of the parallel arms reduces the magnetic moment. It also
shifts the magnetic resonance to higher frequencies. This is a
combined effect of shortened arc length and reduced the geometric
loop inductance. In the limit of only the bottom arm left (i.e.\@ a
straight rod) the magnetic response would vanish (green dotted
curve), and no magnetic response is seen, as expected. In this
limiting case, the resonance in $\mathrm{Re}(\epsilon)$ still
exists, which is a result from the electric resonance of the bottom
arm of SRR.
Note that the magnitude of the resonance in $\mathrm{Re}(\epsilon)$
decreases much slower than the magnetic moment does as the arm
length, $l_y$, decreases. The reason is that the magnetic moment,
$m_z$, resulting from the circular current flowing in the loops of
the SRR, vanishes immediately as the side arms disappear, but the
resonance in $\mathrm{Re}(\epsilon)$ always exist as long as the
short-wire-like electric resonance exists in the bottom part of the
SRR. However, as shown in green curve of Fig. \ref{fig7_Mz}(a), this
short-wire-like electric resonance can not provide any magnetic
moment.

The short-wire-like resonance with the propagating direction of the
incident EM wave perpendicular to the SRR plane and the electirc
field, $\bf E$, parallel to the side arms of the SRR
(Fig.~\ref{fig1_structure}(c.ii)), has zero magnetic moment.

\section{Conclusions}
We systematically studied the electric and magnetic resonances of
U-shaped SRRs with respective to different polarizations of the
incident EM wave. Higher order excitation modes were found in both
electric and magnetic resonances. We show that the magnetic
resonance are the modes with odd-number of half-wavelength of the
current density wave, while the electric resonance are modes with
integer number of whole-wavelength of the current density wave. In
addition, the current density distribution of the lowest three
excitation modes were given. We also studied the magnetization
density as a function of the length of the side arms of the U-shaped
SRRs. It turns out that the magnetic resonance vanishes as the
length of side arms reduces to zero, i.e. a rod does not give any
magnetic moment or magnetic resonance.

\section{Acknowledgments}%
Work at Ames Laboratory was supported by Dept. of Energy (Basic
Energy Sciences) under contract No. DE-AC02-07CH11358, by the AFOSR
under MURI grant (FA9550-06-1-0337), by Dept. of Navy, office of
Naval Research (Award No. N0014-07-1-0359), EU FET projects
Metamorphose and PHOREMOST, and by Greek Ministry of Education
Pythagoras project.


\begin{thebibliography}{27}
\expandafter\ifx\csname
natexlab\endcsname\relax\def\natexlab#1{#1}\fi
\expandafter\ifx\csname bibnamefont\endcsname\relax
  \def\bibnamefont#1{#1}\fi
\expandafter\ifx\csname bibfnamefont\endcsname\relax
  \def\bibfnamefont#1{#1}\fi
\expandafter\ifx\csname citenamefont\endcsname\relax
  \def\citenamefont#1{#1}\fi
\expandafter\ifx\csname url\endcsname\relax
  \def\url#1{\texttt{#1}}\fi
\expandafter\ifx\csname urlprefix\endcsname\relax\def\urlprefix{URL
}\fi \providecommand{\bibinfo}[2]{#2}
\providecommand{\eprint}[2][]{\url{#2}}

\bibitem[{\citenamefont{Veselago}(1968)}]{Veselago_1968}
\bibinfo{author}{\bibfnamefont{V.~G.} \bibnamefont{Veselago}},
  \bibinfo{journal}{Sov. Phys. Usp.} \textbf{\bibinfo{volume}{10}},
  \bibinfo{pages}{509} (\bibinfo{year}{1968}).

\bibitem[{\citenamefont{Smith et~al.}(2004)\citenamefont{Smith, Pendry, and
  Wiltshire}}]{SCI_smith_pendry_review_2004}
\bibinfo{author}{\bibfnamefont{D.~R.} \bibnamefont{Smith}},
  \bibinfo{author}{\bibfnamefont{J.~B.} \bibnamefont{Pendry}},
  \bibnamefont{and} \bibinfo{author}{\bibfnamefont{M.~C.~K.}
  \bibnamefont{Wiltshire}}, \bibinfo{journal}{Science}
  \textbf{\bibinfo{volume}{305}}, \bibinfo{pages}{788} (\bibinfo{year}{2004}).

\bibitem[{\citenamefont{Soukoulis et~al.}(2006)\citenamefont{Soukoulis,
  Kafesaki, and Economou}}]{Adv_Mat_soukoulis_review_2006}
\bibinfo{author}{\bibfnamefont{C.~M.} \bibnamefont{Soukoulis}},
  \bibinfo{author}{\bibfnamefont{M.}~\bibnamefont{Kafesaki}}, \bibnamefont{and}
  \bibinfo{author}{\bibfnamefont{E.~N.} \bibnamefont{Economou}},
  \bibinfo{journal}{Advanced Materials} \textbf{\bibinfo{volume}{18}},
  \bibinfo{pages}{1941} (\bibinfo{year}{2006}).

\bibitem[{\citenamefont{Linden et~al.}(2006)\citenamefont{Linden, Enkrich,
  Dolling, Klein, Zhou, Koschny, Soukoulis, Burger, Schmidt, and
  Wegener}}]{Linden_IEEE_QE_2006}
\bibinfo{author}{\bibfnamefont{S.}~\bibnamefont{Linden}},
  \bibinfo{author}{\bibfnamefont{C.}~\bibnamefont{Enkrich}},
  \bibinfo{author}{\bibfnamefont{G.}~\bibnamefont{Dolling}},
  \bibinfo{author}{\bibfnamefont{M.~W.} \bibnamefont{Klein}},
  \bibinfo{author}{\bibfnamefont{J.~F.} \bibnamefont{Zhou}},
  \bibinfo{author}{\bibfnamefont{T.}~\bibnamefont{Koschny}},
  \bibinfo{author}{\bibfnamefont{C.~M.} \bibnamefont{Soukoulis}},
  \bibinfo{author}{\bibfnamefont{S.}~\bibnamefont{Burger}},
  \bibinfo{author}{\bibfnamefont{F.}~\bibnamefont{Schmidt}}, \bibnamefont{and}
  \bibinfo{author}{\bibfnamefont{M.}~\bibnamefont{Wegener}},
  \bibinfo{journal}{IEEE Journal of Selected Topics in Quantum Electronics}
  \textbf{\bibinfo{volume}{12}}, \bibinfo{pages}{1097} (\bibinfo{year}{2006}).

\bibitem[{\citenamefont{Shalaev}(2007)}]{Nat_Phonotics_shalaev_2007}
\bibinfo{author}{\bibfnamefont{V.~M.} \bibnamefont{Shalaev}},
  \bibinfo{journal}{Nature Photonics} \textbf{\bibinfo{volume}{1}},
  \bibinfo{pages}{41} (\bibinfo{year}{2007}).

\bibitem[{\citenamefont{Soukoulis et~al.}(2007)\citenamefont{Soukoulis, Linden,
  and Wegener}}]{Sci_soukoulis_2007}
\bibinfo{author}{\bibfnamefont{C.~M.} \bibnamefont{Soukoulis}},
  \bibinfo{author}{\bibfnamefont{S.}~\bibnamefont{Linden}}, \bibnamefont{and}
  \bibinfo{author}{\bibfnamefont{M.}~\bibnamefont{Wegener}},
  \bibinfo{journal}{Science} \textbf{\bibinfo{volume}{315}},
  \bibinfo{pages}{47} (\bibinfo{year}{2007}).

\bibitem[{\citenamefont{Pendry et~al.}(1999)\citenamefont{Pendry, Holden,
  Robbins, and Stewart}}]{IEEE_pendry_srr}
\bibinfo{author}{\bibfnamefont{J.}~\bibnamefont{Pendry}},
  \bibinfo{author}{\bibfnamefont{A.}~\bibnamefont{Holden}},
  \bibinfo{author}{\bibfnamefont{D.}~\bibnamefont{Robbins}}, \bibnamefont{and}
  \bibinfo{author}{\bibfnamefont{W.}~\bibnamefont{Stewart}},
  \bibinfo{journal}{IEEE Trans. Microwave Theroy Tech.}
  \textbf{\bibinfo{volume}{47}}, \bibinfo{pages}{2075} (\bibinfo{year}{1999}).

\bibitem[{\citenamefont{Smith et~al.}(2000)\citenamefont{Smith, Padilla, Vier,
  Nemat-Nasser, and Schultz}}]{PRL_Smith_first_NIM}
\bibinfo{author}{\bibfnamefont{D.}~\bibnamefont{Smith}},
  \bibinfo{author}{\bibfnamefont{W.}~\bibnamefont{Padilla}},
  \bibinfo{author}{\bibfnamefont{D.}~\bibnamefont{Vier}},
  \bibinfo{author}{\bibfnamefont{S.}~\bibnamefont{Nemat-Nasser}},
  \bibnamefont{and} \bibinfo{author}{\bibfnamefont{S.}~\bibnamefont{Schultz}},
  \bibinfo{journal}{Physical Review Letters} \textbf{\bibinfo{volume}{84}},
  \bibinfo{pages}{4184} (\bibinfo{year}{2000}).

\bibitem[{\citenamefont{Yen et~al.}(2004)\citenamefont{Yen, Padilla, Fang,
  Vier, Smith, Pendry, Basov, and Zhang}}]{sci_yen_Zhang_Smith_2004}
\bibinfo{author}{\bibfnamefont{T.~J.} \bibnamefont{Yen}},
  \bibinfo{author}{\bibfnamefont{W.~J.} \bibnamefont{Padilla}},
  \bibinfo{author}{\bibfnamefont{N.}~\bibnamefont{Fang}},
  \bibinfo{author}{\bibfnamefont{D.~C.} \bibnamefont{Vier}},
  \bibinfo{author}{\bibfnamefont{D.~R.} \bibnamefont{Smith}},
  \bibinfo{author}{\bibfnamefont{J.~B.} \bibnamefont{Pendry}},
  \bibinfo{author}{\bibfnamefont{D.~N.} \bibnamefont{Basov}}, \bibnamefont{and}
  \bibinfo{author}{\bibfnamefont{X.}~\bibnamefont{Zhang}},
  \bibinfo{journal}{Science} \textbf{\bibinfo{volume}{303}},
  \bibinfo{pages}{1494} (\bibinfo{year}{2004}).

\bibitem[{\citenamefont{Linden et~al.}(2004)\citenamefont{Linden, Enkrich,
  Wegener, Zhou, Koschny, and Soukoulis}}]{science_Linden_100THz}
\bibinfo{author}{\bibfnamefont{S.}~\bibnamefont{Linden}},
  \bibinfo{author}{\bibfnamefont{C.}~\bibnamefont{Enkrich}},
  \bibinfo{author}{\bibfnamefont{M.}~\bibnamefont{Wegener}},
  \bibinfo{author}{\bibfnamefont{J.~F.} \bibnamefont{Zhou}},
  \bibinfo{author}{\bibfnamefont{T.}~\bibnamefont{Koschny}}, \bibnamefont{and}
  \bibinfo{author}{\bibfnamefont{C.~M.} \bibnamefont{Soukoulis}},
  \bibinfo{journal}{Science} \textbf{\bibinfo{volume}{306}},
  \bibinfo{pages}{1351} (\bibinfo{year}{2004}).

\bibitem[{\citenamefont{Enkrich et~al.}(2005)\citenamefont{Enkrich, Wegener,
  Linden, Burger, Zschiedrich, Schmidt, Zhou, Koschny, and
  Soukoulis}}]{PRL_Enkrich_usrr}
\bibinfo{author}{\bibfnamefont{C.}~\bibnamefont{Enkrich}},
  \bibinfo{author}{\bibfnamefont{M.}~\bibnamefont{Wegener}},
  \bibinfo{author}{\bibfnamefont{S.}~\bibnamefont{Linden}},
  \bibinfo{author}{\bibfnamefont{S.}~\bibnamefont{Burger}},
  \bibinfo{author}{\bibfnamefont{L.}~\bibnamefont{Zschiedrich}},
  \bibinfo{author}{\bibfnamefont{F.}~\bibnamefont{Schmidt}},
  \bibinfo{author}{\bibfnamefont{J.~F.} \bibnamefont{Zhou}},
  \bibinfo{author}{\bibfnamefont{T.}~\bibnamefont{Koschny}}, \bibnamefont{and}
  \bibinfo{author}{\bibfnamefont{C.~M.} \bibnamefont{Soukoulis}},
  \bibinfo{journal}{Physical Review Letters} \textbf{\bibinfo{volume}{95}},
  \bibinfo{eid}{203901} (pages~\bibinfo{numpages}{4}) (\bibinfo{year}{2005}),
  \urlprefix\url{http://link.aps.org/abstract/PRL/v95/e203901}.

\bibitem[{\citenamefont{Katsarakis et~al.}(2005)\citenamefont{Katsarakis,
  Konstantinidis, Kostopoulos, Penciu, Gundogdu, Kafesaki, Economou, Koschny,
  and Soukoulis}}]{katsarakis_OL_2005}
\bibinfo{author}{\bibfnamefont{N.}~\bibnamefont{Katsarakis}},
  \bibinfo{author}{\bibfnamefont{G.}~\bibnamefont{Konstantinidis}},
  \bibinfo{author}{\bibfnamefont{A.}~\bibnamefont{Kostopoulos}},
  \bibinfo{author}{\bibfnamefont{R.}~\bibnamefont{Penciu}},
  \bibinfo{author}{\bibfnamefont{T.}~\bibnamefont{Gundogdu}},
  \bibinfo{author}{\bibfnamefont{M.}~\bibnamefont{Kafesaki}},
  \bibinfo{author}{\bibfnamefont{E.}~\bibnamefont{Economou}},
  \bibinfo{author}{\bibfnamefont{T.}~\bibnamefont{Koschny}}, \bibnamefont{and}
  \bibinfo{author}{\bibfnamefont{C.~M.} \bibnamefont{Soukoulis}},
  \bibinfo{journal}{Optics Letters} \textbf{\bibinfo{volume}{30}},
  \bibinfo{pages}{1348} (\bibinfo{year}{2005}).

\bibitem[{\citenamefont{Katsarakis et~al.}(2004)\citenamefont{Katsarakis,
  Koschny, Kafesaki, Economou, and Soukoulis}}]{katsarakis_apl_2004}
\bibinfo{author}{\bibfnamefont{N.}~\bibnamefont{Katsarakis}},
  \bibinfo{author}{\bibfnamefont{T.}~\bibnamefont{Koschny}},
  \bibinfo{author}{\bibfnamefont{M.}~\bibnamefont{Kafesaki}},
  \bibinfo{author}{\bibfnamefont{E.~N.} \bibnamefont{Economou}},
  \bibnamefont{and} \bibinfo{author}{\bibfnamefont{C.~M.}
  \bibnamefont{Soukoulis}}, \bibinfo{journal}{Applied Physics Letters}
  \textbf{\bibinfo{volume}{84}}, \bibinfo{pages}{2943} (\bibinfo{year}{2004}),
  \urlprefix\url{http://link.aip.org/link/?APL/84/2943/1}.

\bibitem[{\citenamefont{Zhou et~al.}(2006{\natexlab{a}})\citenamefont{Zhou,
  Zhang, Tuttle, Koschny, and Soukoulis}}]{PRB_zhou_cwp}
\bibinfo{author}{\bibfnamefont{J.~F.} \bibnamefont{Zhou}},
  \bibinfo{author}{\bibfnamefont{L.}~\bibnamefont{Zhang}},
  \bibinfo{author}{\bibfnamefont{G.}~\bibnamefont{Tuttle}},
  \bibinfo{author}{\bibfnamefont{T.}~\bibnamefont{Koschny}}, \bibnamefont{and}
  \bibinfo{author}{\bibfnamefont{C.~M.} \bibnamefont{Soukoulis}},
  \bibinfo{journal}{Physical Review B (Condensed Matter and Materials Physics)}
  \textbf{\bibinfo{volume}{73}}, \bibinfo{eid}{041101}
  (pages~\bibinfo{numpages}{4}) (\bibinfo{year}{2006}{\natexlab{a}}),
  \urlprefix\url{http://link.aps.org/abstract/PRB/v73/e041101}.

\bibitem[{\citenamefont{Zhou et~al.}(2006{\natexlab{b}})\citenamefont{Zhou,
  Koschny, Zhang, Tuttle, and Soukoulis}}]{APL_zhou_cwp}
\bibinfo{author}{\bibfnamefont{J.~F.} \bibnamefont{Zhou}},
  \bibinfo{author}{\bibfnamefont{T.}~\bibnamefont{Koschny}},
  \bibinfo{author}{\bibfnamefont{L.}~\bibnamefont{Zhang}},
  \bibinfo{author}{\bibfnamefont{G.}~\bibnamefont{Tuttle}}, \bibnamefont{and}
  \bibinfo{author}{\bibfnamefont{C.~M.} \bibnamefont{Soukoulis}},
  \bibinfo{journal}{Applied Physics Letters} \textbf{\bibinfo{volume}{88}},
  \bibinfo{eid}{221103} (pages~\bibinfo{numpages}{3})
  (\bibinfo{year}{2006}{\natexlab{b}}),
  \urlprefix\url{http://link.aip.org/link/?APL/88/221103/1}.

\bibitem[{\citenamefont{Zhang et~al.}(2005{\natexlab{a}})\citenamefont{Zhang,
  Fan, Minhas, Frauenglass, Malloy, and Brueck}}]{PRL_zhang_2005_037402}
\bibinfo{author}{\bibfnamefont{S.}~\bibnamefont{Zhang}},
  \bibinfo{author}{\bibfnamefont{W.}~\bibnamefont{Fan}},
  \bibinfo{author}{\bibfnamefont{B.~K.} \bibnamefont{Minhas}},
  \bibinfo{author}{\bibfnamefont{A.}~\bibnamefont{Frauenglass}},
  \bibinfo{author}{\bibfnamefont{K.~J.} \bibnamefont{Malloy}},
  \bibnamefont{and} \bibinfo{author}{\bibfnamefont{S.~R.~J.}
  \bibnamefont{Brueck}}, \bibinfo{journal}{Physical Review Letters}
  \textbf{\bibinfo{volume}{94}}, \bibinfo{eid}{037402}
  (pages~\bibinfo{numpages}{4}) (\bibinfo{year}{2005}{\natexlab{a}}),
  \urlprefix\url{http://link.aps.org/abstract/PRL/v94/e037402}.

\bibitem[{\citenamefont{Zhang et~al.}(2005{\natexlab{b}})\citenamefont{Zhang,
  Fan, Panoiu, Malloy, Osgood, and Brueck}}]{PRL_zhang_2005_137404}
\bibinfo{author}{\bibfnamefont{S.}~\bibnamefont{Zhang}},
  \bibinfo{author}{\bibfnamefont{W.}~\bibnamefont{Fan}},
  \bibinfo{author}{\bibfnamefont{N.~C.} \bibnamefont{Panoiu}},
  \bibinfo{author}{\bibfnamefont{K.~J.} \bibnamefont{Malloy}},
  \bibinfo{author}{\bibfnamefont{R.~M.} \bibnamefont{Osgood}},
  \bibnamefont{and} \bibinfo{author}{\bibfnamefont{S.~R.~J.}
  \bibnamefont{Brueck}}, \bibinfo{journal}{Physical Review Letters}
  \textbf{\bibinfo{volume}{95}}, \bibinfo{eid}{137404}
  (pages~\bibinfo{numpages}{4}) (\bibinfo{year}{2005}{\natexlab{b}}),
  \urlprefix\url{http://link.aps.org/abstract/PRL/v95/e137404}.

\bibitem[{\citenamefont{Dolling
  et~al.}(2006{\natexlab{a}})\citenamefont{Dolling, Enkrich, Wegener,
  Soukoulis, and Linden}}]{dolling_OL_2006}
\bibinfo{author}{\bibfnamefont{G.}~\bibnamefont{Dolling}},
  \bibinfo{author}{\bibfnamefont{C.}~\bibnamefont{Enkrich}},
  \bibinfo{author}{\bibfnamefont{M.}~\bibnamefont{Wegener}},
  \bibinfo{author}{\bibfnamefont{C.~M.} \bibnamefont{Soukoulis}},
  \bibnamefont{and} \bibinfo{author}{\bibfnamefont{S.}~\bibnamefont{Linden}},
  \bibinfo{journal}{Optics Letters} \textbf{\bibinfo{volume}{31}},
  \bibinfo{pages}{1800} (\bibinfo{year}{2006}{\natexlab{a}}).

\bibitem[{\citenamefont{Dolling
  et~al.}(2006{\natexlab{b}})\citenamefont{Dolling, Enkrich, Wegener,
  Soukoulis, and Linden}}]{science_soukoulis_2006}
\bibinfo{author}{\bibfnamefont{G.}~\bibnamefont{Dolling}},
  \bibinfo{author}{\bibfnamefont{C.}~\bibnamefont{Enkrich}},
  \bibinfo{author}{\bibfnamefont{M.}~\bibnamefont{Wegener}},
  \bibinfo{author}{\bibfnamefont{C.~M.} \bibnamefont{Soukoulis}},
  \bibnamefont{and} \bibinfo{author}{\bibfnamefont{S.}~\bibnamefont{Linden}},
  \bibinfo{journal}{Science} \textbf{\bibinfo{volume}{312}},
  \bibinfo{pages}{892} (\bibinfo{year}{2006}{\natexlab{b}}).

\bibitem[{\citenamefont{Dolling et~al.}(2007)\citenamefont{Dolling, Wegener,
  Soukoulis, and Linden}}]{dolling_fishnet_780nm_2007}
\bibinfo{author}{\bibfnamefont{G.}~\bibnamefont{Dolling}},
  \bibinfo{author}{\bibfnamefont{M.}~\bibnamefont{Wegener}},
  \bibinfo{author}{\bibfnamefont{C.~M.} \bibnamefont{Soukoulis}},
  \bibnamefont{and} \bibinfo{author}{\bibfnamefont{S.}~\bibnamefont{Linden}},
  \bibinfo{journal}{Optics Lett.} \textbf{\bibinfo{volume}{32}},
  \bibinfo{pages}{53} (\bibinfo{year}{2007}).

\bibitem[{\citenamefont{Chettiar et~al.}(2007)\citenamefont{Chettiar,
  Kildishev, Yuan, Cai, Xiao, Drachev, and Shalaev}}]{Opl_Chettiar_2007}
\bibinfo{author}{\bibfnamefont{U.~K.} \bibnamefont{Chettiar}},
  \bibinfo{author}{\bibfnamefont{A.~V.} \bibnamefont{Kildishev}},
  \bibinfo{author}{\bibfnamefont{H.~K.} \bibnamefont{Yuan}},
  \bibinfo{author}{\bibfnamefont{W.}~\bibnamefont{Cai}},
  \bibinfo{author}{\bibfnamefont{S.}~\bibnamefont{Xiao}},
  \bibinfo{author}{\bibfnamefont{V.~P.} \bibnamefont{Drachev}},
  \bibnamefont{and} \bibinfo{author}{\bibfnamefont{V.~M.}
  \bibnamefont{Shalaev}}, \bibinfo{journal}{Optics Letters}
  \textbf{\bibinfo{volume}{32}}, \bibinfo{pages}{1671} (\bibinfo{year}{2007}),
  \urlprefix\url{http://www.opticsinfobase.org/abstract.cfm?URI=ol-32-12-1671}.

\bibitem[{\citenamefont{Smith et~al.}(2002)\citenamefont{Smith, Schultz,
  Markos, and Soukoulis}}]{PRB_smith_retrieval_2002}
\bibinfo{author}{\bibfnamefont{D.~R.} \bibnamefont{Smith}},
  \bibinfo{author}{\bibfnamefont{S.}~\bibnamefont{Schultz}},
  \bibinfo{author}{\bibfnamefont{P.}~\bibnamefont{Markos}}, \bibnamefont{and}
  \bibinfo{author}{\bibfnamefont{C.~M.} \bibnamefont{Soukoulis}},
  \bibinfo{journal}{Physical Review B (Condensed Matter and Materials Physics)}
  \textbf{\bibinfo{volume}{65}}, \bibinfo{eid}{195104}
  (pages~\bibinfo{numpages}{5}) (\bibinfo{year}{2002}),
  \urlprefix\url{http://link.aps.org/abstract/PRB/v65/e195104}.

\bibitem[{\citenamefont{Smith et~al.}(2005)\citenamefont{Smith, Vier, Koschny,
  and Soukoulis}}]{PRE_Smith_retrieval_2005}
\bibinfo{author}{\bibfnamefont{D.~R.} \bibnamefont{Smith}},
  \bibinfo{author}{\bibfnamefont{D.~C.} \bibnamefont{Vier}},
  \bibinfo{author}{\bibfnamefont{T.}~\bibnamefont{Koschny}}, \bibnamefont{and}
  \bibinfo{author}{\bibfnamefont{C.~M.} \bibnamefont{Soukoulis}},
  \bibinfo{journal}{Phys. Rev. E} \textbf{\bibinfo{volume}{71}},
  \bibinfo{pages}{036617} (\bibinfo{year}{2005}).

\bibitem[{\citenamefont{Koschny et~al.}(2005)\citenamefont{Koschny, Markos,
  Economou, Smith, Vier, and Soukoulis}}]{PRB_koschny_retrieval_2005}
\bibinfo{author}{\bibfnamefont{T.}~\bibnamefont{Koschny}},
  \bibinfo{author}{\bibfnamefont{P.}~\bibnamefont{Markos}},
  \bibinfo{author}{\bibfnamefont{E.~N.} \bibnamefont{Economou}},
  \bibinfo{author}{\bibfnamefont{D.~R.} \bibnamefont{Smith}},
  \bibinfo{author}{\bibfnamefont{D.~C.} \bibnamefont{Vier}}, \bibnamefont{and}
  \bibinfo{author}{\bibfnamefont{C.~M.} \bibnamefont{Soukoulis}},
  \bibinfo{journal}{Physical Review B (Condensed Matter and Materials Physics)}
  \textbf{\bibinfo{volume}{71}}, \bibinfo{eid}{245105}
  (pages~\bibinfo{numpages}{22}) (\bibinfo{year}{2005}),
  \urlprefix\url{http://link.aps.org/abstract/PRB/v71/e245105}.

\bibitem[{\citenamefont{Koschny et~al.}(2004)\citenamefont{Koschny, Kafesaki,
  Economou, and Soukoulis}}]{PRL_koschny_2004}
\bibinfo{author}{\bibfnamefont{T.}~\bibnamefont{Koschny}},
  \bibinfo{author}{\bibfnamefont{M.}~\bibnamefont{Kafesaki}},
  \bibinfo{author}{\bibfnamefont{E.~N.} \bibnamefont{Economou}},
  \bibnamefont{and} \bibinfo{author}{\bibfnamefont{C.~M.}
  \bibnamefont{Soukoulis}}, \bibinfo{journal}{Phys. Rev. Lett.}
  \textbf{\bibinfo{volume}{93}} (\bibinfo{year}{2004}).

\bibitem[{\citenamefont{Koschny et~al.}(2003)\citenamefont{Koschny, Markos,
  Smith, and Soukoulis}}]{PRE_koschny_retrieval_2003}
\bibinfo{author}{\bibfnamefont{T.}~\bibnamefont{Koschny}},
  \bibinfo{author}{\bibfnamefont{P.}~\bibnamefont{Markos}},
  \bibinfo{author}{\bibfnamefont{D.~R.} \bibnamefont{Smith}}, \bibnamefont{and}
  \bibinfo{author}{\bibfnamefont{C.~M.} \bibnamefont{Soukoulis}},
  \bibinfo{journal}{Phys. Rev. E} \textbf{\bibinfo{volume}{68}},
  \bibinfo{pages}{065602} (\bibinfo{year}{2003}).

\bibitem[{\citenamefont{Rockstuhl et~al.}(2006)\citenamefont{Rockstuhl,
  Lederer, Etrich, Zentgraf, Kuhl, and Giessen}}]{opl_rockstuhl_2006}
\bibinfo{author}{\bibfnamefont{C.}~\bibnamefont{Rockstuhl}},
  \bibinfo{author}{\bibfnamefont{F.}~\bibnamefont{Lederer}},
  \bibinfo{author}{\bibfnamefont{C.}~\bibnamefont{Etrich}},
  \bibinfo{author}{\bibfnamefont{T.}~\bibnamefont{Zentgraf}},
  \bibinfo{author}{\bibfnamefont{J.}~\bibnamefont{Kuhl}}, \bibnamefont{and}
  \bibinfo{author}{\bibfnamefont{H.}~\bibnamefont{Giessen}},
  \bibinfo{journal}{Optics Express} \textbf{\bibinfo{volume}{14}},
  \bibinfo{pages}{8827} (\bibinfo{year}{2006}).

\end{thebibliography}

\end{document}